\documentclass[aps,prd]{revtex4}

\usepackage{amssymb}
\usepackage{hyperref}

\usepackage{graphicx} 
\usepackage[scriptsize,hang,flushleft]{caption}
\usepackage[scriptsize,hang]{subfigure}

\usepackage{color}
\usepackage{placeins}

\usepackage{amsmath,amsfonts,bbm,dsfont,mathrsfs,txfonts}

\usepackage[T1]{fontenc}
\usepackage[ngerman, english]{babel} 
\usepackage[utf8]{inputenc}

\newcommand{\half}{\frac{1}{2}}
\newcommand{\dd}{{\mathrm{d}}}

\begin{document}
\title{The ISCO of charged particles in Reissner-Nordström, Kerr-Newman and Kerr-Sen spacetime}

\author{Kris Schroven}
\email[]{kris.schroven@asu.cas.cz}
\affiliation{Astronomical Institute, Czech Academy of Sciences}
\author{Saskia Grunau}
\email[]{saskia.grunau@uni-oldenburg.de}
\affiliation{Institut für Physik, Universität Oldenburg, D-26111 Oldenburg, Germany}

\date{\today}

\begin{abstract}
In this article we study the innermost stable circular orbit (ISCO) of electrically charged particles in the electrically charged Reissner-Nordström spacetime, the Kerr-Newman spacetime and the Kerr-Sen spacetime. We find that the radius of the ISCO increases with an increasing particle-black hole charge product $|qQ|$ in the case of attractive Coulomb interaction $qQ<0$. For repulsive Coulomb interaction, the ISCO radius first decreases to a minimum and then increases again, until it diverges as the charge product approaches one.
If the charge $Q$ of the black hole is very small, the minimum of the ISCO radius lies at $qQ=0$. Repulsive and attractive Coulomb interactions will always increase the ISCO radius in this limit.
Stable bound orbits of charged particles cease to exist in the Reissner-Nordström and Kerr-Newman spacetime  for $qQ\geq 1$. In the Kerr-Sen spacetime the limiting case  depends on the charge of the black hole and if dilaton coupling is applied to the test particle. We find $qQ\geq 1+Q^2$ without dilaton-coupling and $qQ\geq 1+\frac{3}{2}Q^2$ with dilaton coupling $\alpha=1$.
\end{abstract}

\maketitle
\section{Introduction}

The dynamics of particles in the vicinity of black holes in relativistic astrophysics exhibits many interesting phenomena. One of the relativistic effects is the existence of an innermost stable circular orbit (ISCO), which represents the boundary between test particles orbiting the black hole and test particles falling into the black hole. It is therefore an important feature for accretion disk physics, since it marks the inner edge of the accretion disk in the thin disk model of Shakura and Sunjaev \cite{Shakura:1972te, Abramowicz:2011xu}.
It is further used in the thick disc model as a limit for the parameter space that yields to bound solutions \cite{Abramowicz:2011xu, Abramowicz:1978}. In accretion disk simulations, which can be compared to EHT observations, these models are often used as a starting point \cite{Akiyama:2019cqa}.

In this article we are interested in the ISCO of charged particles in charged black hole spacetimes. When dealing with the orbits of charged particles, the charge of the black hole is not neglible. However, the charge of real black holes is expected to be very small. The charge of the central black hole of our galaxy, Sgr A$^*$, was recently estimated to be at most $3\times 10^8$C (or $4\times 10^{-19}M$) in terms of the black hole mass) \cite{Zajacek:2018ycb}. Such a small charge of a black hole will problably not influence the curvature around it, but it will have a significant effect on the ISCO of charged particles \cite{Zajacek:2018ycb, Zajacek:2019kla}. An observation of the ISCO by e.g. X-ray radiation could be used to get information both on the rotation and the charge of a black hole.

In \cite{Pugliese:2011py} the circular motion of electrically charged test particles in the electrically charged Reissner-Nordström spacetime was analysed in detail. Furthermore the ISCO is studied and an equation for the angular momentum for a charged particle on the ISCO was given. The authors found that in general the radius of the ISCO increases with increasing charge $|q|$ of the test particle. In the case of attractive Coulomb interaction $qQ<0$, the Coulomb force reinforces the gravitational interaction and charged particles behave similar to neutral test particles. In the case of repulsive Coulomb interaction $qQ>0$ the situation is more complicated and for a certain parameter region of particle and black hole charges stable bound orbits do not exist at all.

The motion of charged test particles in Kerr-Newman spacetime was studied in a series of paper in \cite{Bicak,Balek}. Finally, the analytical solution was presented in terms of Weierstass elliptic functions by \cite{Hackmann:2013pva}. The ISCO of electrically charged particles in the electrically charged Kerr-Newman spacetime were studied in \cite{Schroven:2017jsp}. Here the charge $Q$ of the black hole was estimated to be very small, so that only interaction terms, that contain the product of particle and black hole charge $qQ$, would enter the equations. The minimal ISCO radius in that case is found for uncharged particles $qQ=0$, identifying the ISCOs in Kerr spacetime as the lower limit for ISCO radii of charged particles at a given spin. The ISCO radius grows with increasing $|qQ|$, however, ISCOs cease to exist for $qQ=1$.

		A strong electromagnetic field caused by a black hole charge evokes electron-positron pair creation. Selective accretion will consequently reduce the black hole's charge until $Q \lesssim 10^{-5}$, when the pair creation process will stop \cite{Hanni1982}. Selective accretion of surrounding matter (interstellar medium, etc.) will further reduce the black hole charge until the electromagnetic interaction is comparable to gravitational effects \cite{Zajacek:2018ycb, Eardley:1975kp}. This would imply $Q\sim 10^{-21}$ in the case of electrons as test particles ($q_{\rm electron}\sim 10^{21}$) \cite{Hackstein:2019msh}.  Even though black hole charges strong enough to contribute to spacetime curvature are unlikely on long timescales, they might occur  for some time period after their creation.
		
Since the charge of an astrophysical black hole will probably be very small, one could also imagine a black hole spacetime immersed in an electromagnetic field which will not influence the metric. ISCOs of charged particles around a Schwarzschild black hole in the presence electromagnetic fields were investigated in \cite{Hackstein:2019msh}. It was observed that an electric  field increases the ISCO radius, while a magnetic field decreases the ISCO radius. If the electric field is sufficiently strong and has the same sign as the charge of the particle, then stable bound orbits cannot exist. Here the limiting case is $qQ=1$ as for charged particles in the Reissner-Nordström and Kerr-Newman spacetime. This effect can be cancelled with a sufficiently strong magnetic field so that ISCOs of static particles appear.

The ISCO of charged particles in the spacetime of a quasi-Kerr compact object immersed in a uniform magnetic field was considered numerically in \cite{Narzilloev:2019hep}. As in the Schwarzschild case, an increasing magnetic field will decrease the ISCO radius.\\

In this article we will study the ISCO of electrically charged particles in the electrically charged Reissner-Nordström spacetime, the Kerr-Newman spacetime and the Kerr-Sen spacetime. For mathematical curiosity we will also analyse the region behind the event horizon.

Our study of the Reissner-Nordström spacetime confirms the result of \cite{Pugliese:2011py} and presents new details. In the Kerr-Newman spacetime we will extend the analysis of the ISCO in \cite{Schroven:2017jsp} to arbitrary values of the charge $Q$ of the black hole. 

Another interesting charged rotating black hole is the Kerr-Sen solution \cite{Sen:1992ua}, which arises from four-dimensional heterotic string theory.  Recently, the Kerr-Sen black hole was compared to EHT observations of M87$^*$ \cite{Narang:2020bgo}. We will consider the ISCO of charged particles in the Kerr-Sen spacetime and compare our results to the Reissner-Nordström and Kerr-Newman spacetime.

	\section{The innermost stable circular orbit}
	\label{secI}
		The existence of an innermost stable circular orbit is a purely relativistic effect. In classical mechanics, circular orbits of neutral test particles around any central spherical mass distribution are always stable. And these circular orbits can be arbitrarily close to the central mass.  
		In Newtonian physics, gravitational and electrostatic interactions can both be described by a respective potential, determined by Poisson's equation. Hence, the equations of motions describing the test particle motion in a gravitational field will not change qualitatively, when adding an electric charge to the test particle and central mass.  Again, circular orbits are always stable and can be found arbitrarily close to the central mass, even if test particle and central mass are charged. For attractive electromagnetic interaction as well as repulsive interaction (as long as the repulsive force is smaller than the gravitational force) one can always find an angular momentum, for which bound orbits are possible.
	
		In general relativity - however - massive test particle velocities cannot be equal or exceeding the speed of light. This is given credit to in the equations of motion for a test particle around a Schwarzschild black hole. An additional term $2L^2/r^3$ arises next to the gravitational ($1/r$) and centrifugal ($L^2/r^2$) term in the effective potential $V_{eff}$ of the radial equation of motion:
		\begin{equation}
			\dot r^2=E^2-1-2V_{eff}(r)=E^2-1-2\left(-\frac{1}{r}+\frac{L^2}{2r^2}-\frac{L^2}{r^3}\right)\, .
			\label{SSEOM}
		\end{equation}
		The test particle energy and angular momentum are declared as $E$ and $L$.
		Instead of only one minimum, $V_{eff}$ can now develop either a local maximum and minimum or no extremum, depending on the choice of the angular momentum $L$. The extrema correspond to a stable outer and unstable inner circular test particle orbit. The innermost stable circular orbit occurs for a certain parameter $L$, for which stable and unstable circular orbit merge.
	
		By using the Hamilton-Jacobi formalism, the electromagnetic interaction of a charged test particle and charged central black hole enters the equations of motion as follows:
		\begin{align}
			E-q A_t&=E-V_{el}=-u_t\, , &   L-qA_\phi&= u_\phi\, .
		\end{align}
		$A_\mu$, $\mu=t, \phi$ is the electromagnetic potential of the charged central black hole acting on the charged test particle.
	
		In contrary to the gravitational interaction, the electric one enters the equations analogue to the classical case ($E-V_{el}=\frac{1}{2}v^2+V_{grav}$, with particle velocity $v$). Due to the normalization condition of the velocity in GR, $u_t$ is determined by
		\begin{equation}
			-1=g^{tt} u_t^2+2\,g^{t\phi}u_t u_{\phi}+g^{ab}u_a u_b\, , \text{ for $a,b$ spacial coordinates, eg. $r,\phi$}\, . 
		\end{equation} 
		It will therefore appear quadratically in the equations of motions for all spacial components $u^{a,b}$. In a general relativistic treatment, the electromagnetic potential enters the radial equations of motion not only linearly - like in the classical case - but also quadratically. The  "relativistic" term $A_t^2$ does not distinguish between attractive and repulsive electric forces acting on the charged test particle. 
		
		This is an attempt to understand the growth of the ISCO radius for both an attractive ($qQ<0$) and repulsive ($qQ>0$) electric force on the charged ($q$) test particle for increasing values of the particle-black hole charge product $|qQ|$. We will see this behaviour of the ISCO not only for a Reissner-Nordström and Kerr-Newman, but also for a Kerr-Sen black hole.
		
	\section{ISCO in Reissner-Nordström spacetime}
	The ISCO of charged particles in the Reissner-Nordström spacetime was considered before in \cite{Pugliese:2011py}. We confirm their results, but also present new details.
	
		The Reissner-Nordström metric of an electrically charged black hole is \cite{Reissner:1916, Nordstrom:1918}
		\begin{equation}
		\dd s^2 = -\frac{\Delta}{r^2} \dd t^2 + \frac{r^2}{\Delta} \dd r^2 + r^2\dd\theta ^2+ r^2\sin\theta^2 \dd\phi^2
		\end{equation}
		where $\Delta = r^2-2Mr+Q^2$ and the non-vanishing part of the electromagnetic vector potential is $A_t=\frac{Q}{r}$. The two horizons $r_\pm$ are determined by $\Delta=0$ and exist if $Q^2\leq M^2$
		\begin{equation}
		r_\pm = M\pm\sqrt{M^2-Q^2}\, .
		\end{equation}
		The Hamilton-Jacobi equation for electrically charged particles is
		\begin{equation}
		\frac{\partial S}{\partial \lambda} + \half g^{\mu\nu} \left( \frac{\partial S}{\partial x^\mu} + qA_\mu \right) \left( \frac{\partial S}{\partial x^\nu} + qA_\nu \right) = 0
		\label{eqn:HJD}
		\end{equation}
		where $q$ is the charge of the test particle and $\lambda $ is an affine parameter along the geodesic. Due to spherical symmetry we can restrict the motion to the equatorial plane $\theta = \frac{\pi}{2}$. The Hamilton-Jacobi equation can be solved with the following ansatz for the action
		\begin{equation}
		S= \half\delta\lambda -Et + L\phi + W(r) .
		\end{equation}
		Here $\delta$ is equal to $0$ for light and equal to $1$ for particles. $E$ is the conserved energy and $L$ is the conserved angular momentum of the test particle. Then we can derive the equations of motion from the Hamilton-Jacobi equation
		\begin{align}
		\left( \frac{\dd r}{\dd\phi} \right)^2 &=\frac{r^4}{L^2}\left[\left( E-\frac{qQ}{r}\right)^2 - \frac{\Delta}{r^2} \left( \delta + \frac{L^2}{r^2}\right) \right] = R(r) \, ,\\
		\left( \frac{\dd r}{\dd t }\right)^2 &= \frac{\Delta^2}{r^4} -\frac{\Delta^3}{r^6} \left( \delta + \frac{L^2}{r^2}\right) \left( E-\frac{qQ}{r}\right)^{-2} \, . 
		\label{RNradialEOM}
		\end{align}
		We used scaled quantities in the equations of motion
		\begin{equation}
		r\rightarrow \frac{r}{M} \, , \ \phi\rightarrow \frac{\phi}{M} \, , \ 
		Q\rightarrow \frac{Q}{M} \, , \ L\rightarrow \frac{L}{M} \, .
		\end{equation}
		Note that $R=\sum_{i=0}^4 a_i r^i$ is a polynomial of order 4 with the coefficients.
		\begin{align}
		a_4 &= \frac{1}{L^2} \left( E^2-\delta \right) \, ,\\
		a_3 &=\frac{1}{L^2} \left( -EqQ + \delta \right)\, ,\\
		a_2 &=\frac{1}{L^2} \left( Q^2(q^2-\delta)-L^2\right) \, ,\\
		a_1 &= 2 \, ,\\
		a_0 &=-Q^2\, .
		\end{align}
		The zeros of $\left( \frac{\dd r}{\dd\phi} \right)^2=R$ are the turning points of the geodesics. The number of zeros is related to the possible types of orbits. If $R$ possesses 4 zeros, then there are many-world bound orbits crossing the horizons and bound orbits with turning points $r_{1,2}>r_+$, compare \cite{Grunau:2010gd}. Here we are interested in the latter and especially in the ISCO of charged particles.
		
		Descartes' rule of signs states that the number of sign changes of the coefficients of a polynomial is equal to the number of positive real roots or less by an even number. Therefore, $4$ positive zeros of $R$ can exist if $a_4 <0$, $a_3>0$, $a_2<0$,  $a_1>0$, $a_0<0$. From $a_4 <0$ we can deduce that stable bound orbits with two turning points exist for $\delta=1$, but not for $\delta=0$. Furthermore, we get the following conditions for bound orbits in the equatorial plane and therefore also for ISCOs
		\begin{align}
		\label{RNlimBO1}
		E^2 &< 1 \, ,\\
		qQ &< 1 \, ,\\
		L^2 &> 1-Q^2 \, .
		\label{RNlimBO2}
		\end{align}
		These conditions apply for bound orbits with $r>r_+$, however, in the Reissner-Nordström spacetime bound orbits of charged particles can also exist behind the inner horizon $r<r_-$ \cite{Grunau:2010gd}.
		
		The radial equation of motion (see Eq. \eqref{RNradialEOM}) can be rewritten for $\frac{\dd r}{\dd t }=0$ as
		\begin{equation}
		0 =(\hat E^2-1) +-2\left( -\frac{1}{r} +\frac{\hat{L}^2}{r^2}-\left(1-\frac{Q^2}{2r} \right)\left(\hat{L}^2+\frac{\bar q^2-Q^2}{1-E \,\bar q}\right)\frac{1}{r^3}\right)\, ,
		\end{equation}
		with $\hat L^2=L^2-\frac{\bar q^2-Q^2}{1-E \,\bar q}$ and $\hat E^2 -1= \frac{(E^2-1)}{1-E\,\bar q}$. The charge product of test particle and black hole charge is now declared as $qQ=\bar q$.
		A qualitative comparison with Eq. \eqref{SSEOM} shows, that only the "relativistic" term (originally $\propto 1/r^3$ in Eq. \eqref{SSEOM}) deviates in its structure from the one in the Schwarzschild equation of motion. The bigger the term becomes, its influence on the course of the effective potential $V_{eff}$ grows, and the maximum -- and with it the radius, where minimum and maximum are to merge -- moves to bigger radii.
		
		One can easily see, that for very small charges of the central black hole ($Q\rightarrow 0$), but $\bar q\neq0$, the "relativistic" term is smallest for uncharged test particles ($\bar q=0$), since $\frac{\bar q^2}{1-E \,\bar q}$ is positive for all $E, \bar q$, that can occur for bound orbits (see Eq. \ref{RNlimBO1}-\ref{RNlimBO2}). For  $Q\neq0$, the "relativistic" term will be smaller than in the uncharged case, if $\bar q^2$ is sufficiently small. Hence, the ISCO has to reach its smallest radius for uncharged particles ($\bar q=0$), when the effect of the black hole charge on the spacetime is negligible ($Q\approx0$). However, if the black hole charge significantly effects the spacetime curvature, the ISCO can become smaller than in the uncharged case for a repulsive electromagnetic force ($E\bar q>0$), if $|\bar q|$ is  sufficiently small.
		
\begin{figure}[h]
			\centering
			\subfigure[]{
				\includegraphics[width=0.44\linewidth]{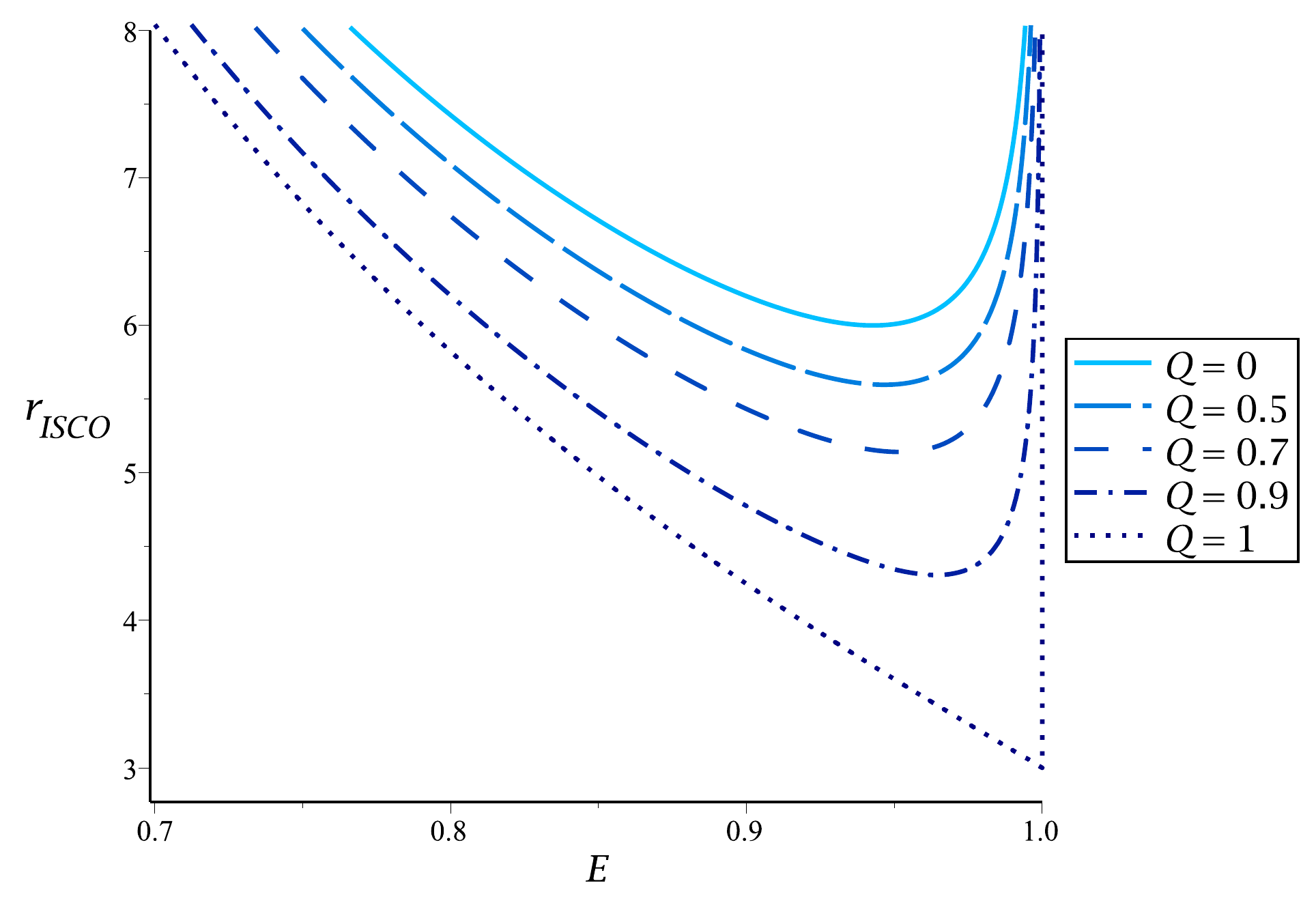}
			}
			\subfigure[]{
				\includegraphics[width=0.44\linewidth]{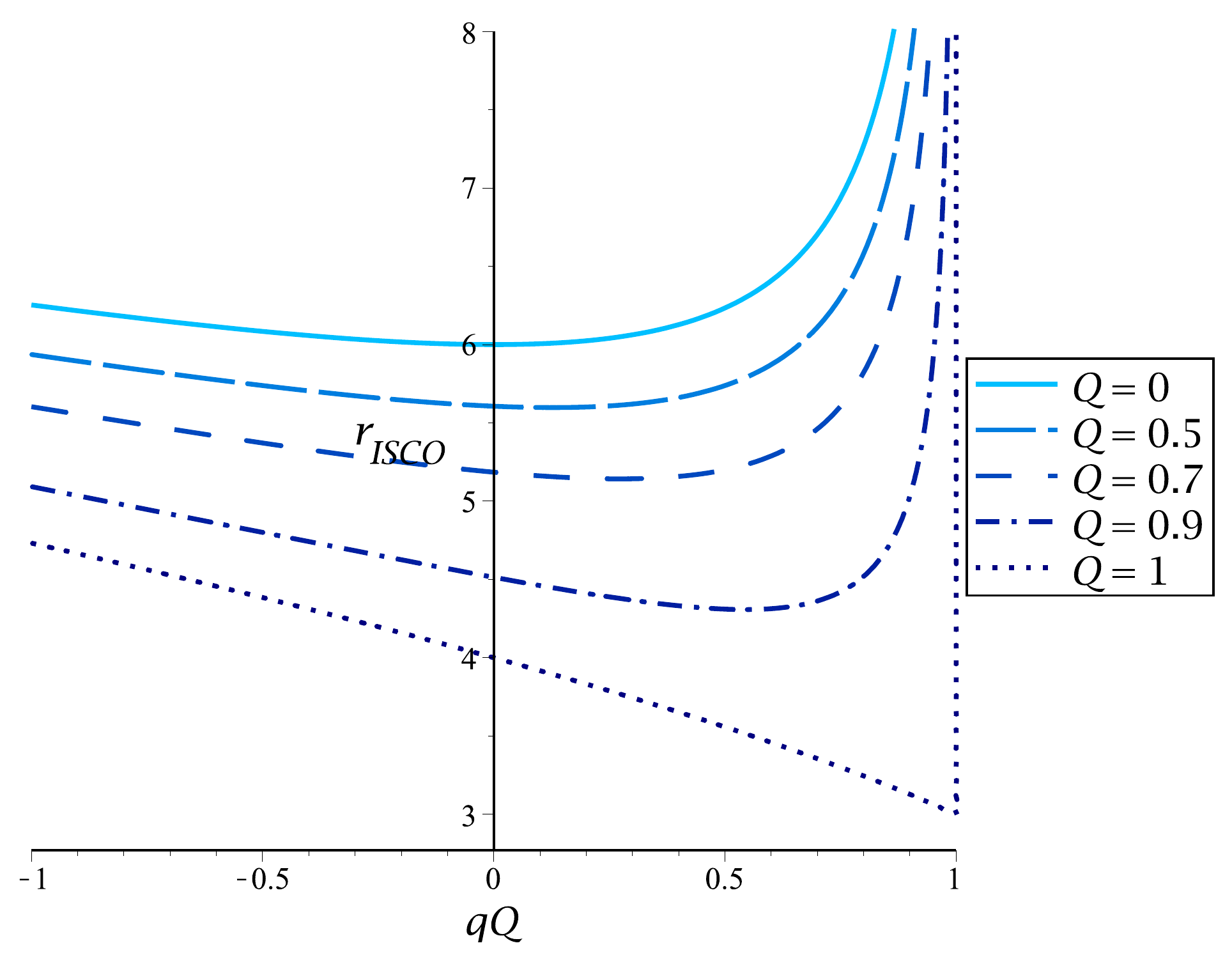}
			}
			\caption{ISCO of electrically charged particles in the Reissner-Nordström spacetime. (a) $r_{\rm ISCO}$ over $E$ for several values of $Q$. (b)  $r_{\rm ISCO}$ over $qQ$ for several values of $Q$.}
			\label{pic:ISCO}
		\end{figure}

		One can define an effective potential by
		\begin{equation}
		R(r)=\frac{r^4}{L^2}(E-V_+)(E-V_-)
		\end{equation}
		so that
		\begin{align}
		V_\pm &= \frac{\bar q}{r}\pm\frac{1}{r^2}\sqrt{\Delta\left( L^2+\delta r^2 \right)}\\
		&= \frac{\bar q}{r}\pm \sqrt{ \delta - \frac{2\delta}{r} + \frac{L^2}{r^2}  - \frac{2L^2}{r^3} + \frac{\delta Q^2}{r^2} + \frac{L^2Q^2}{r^4} }\, . \nonumber
		\end{align}
		The ISCO is located at an inflection point of the effective potential. To calculate the ISCO, three conditions have to be taken into account
		\begin{align}
		R&=0 \, , \label{eqn:cond1}\\ 
		\frac{\dd R}{\dd r}&=0 \, , \label{eqn:cond2}\\ 
		\frac{\dd^2 R}{\dd r^2}&=0 \, . \label{eqn:cond3}
		\end{align}
		We solve \eqref{eqn:cond3} for $L^2$, then we substitute this into \eqref{eqn:cond2} and solve for $q$. With these results we can rewrite \eqref{eqn:cond1} to obtain a condition for the ISCO depending on the location $r$ the energy $E$ and the charge $Q$
		\begin{align}
		&E\left(3Q^2-2r\right) \sqrt{9E^2r^4 -4 \left(5E^2+4\right)r^3 + 12\left(E^2+3\right)r^2-24r+4Q^2}  \nonumber\\
		&+ 6E^2r^3 - 3\left(2+\left(3Q^2+2\right)E^2\right)r^2 + 2\left(5E^2Q^2+4Q^2+2\right)r-6Q^2 =0 \, .
		\label{eq:EISCO-cond}
		\end{align}
		We can plot this equation for different values of the black hole charge $Q$ to obtain a curve for the energy and the location of the ISCO. Figure \ref{pic:ISCO}(a) shows $r_{\rm ISCO}$ over $E$ for several values of $Q$.
		
		Another possibility is to solve \eqref{eqn:cond3} for $L^2$, then substitute this into \eqref{eqn:cond2} and solve for $E$. With these results we can rewrite \eqref{eqn:cond1} to obtain a condition for the ISCO depending on the location $r$ the charge of the particle $q$ and the charge of the black hole $Q$
		\begin{align}
		&\bar{q} \left( 6Q^2+r^2-6r \right) \sqrt{64r^4-240r^3 + 3\left( 3\bar{q}^2+104 \right) r^2 -4 \left(4Q^2+5\bar{q}^2+36 \right) r +24Q^2+12\bar{q}^2} \nonumber \\ 
		&-8r^4 + 3\left( \bar{q}^2+20 \right) r^3 -72 \left( Q^2 +1 \right) r^2 + 2\left( 16Q^4+ \left( 54-7\bar{q}^2 \right) Q^2 \right) r-48Q^4+12Q^2\bar{q}^2 =0\, .
		\label{eq:rISCO-cond}
		\end{align}
		
		The ISCO in RN spacetime will approach infinity, when the prefactor of the three monomials of $R(r)$ with the three highest degrees in $r$  vanish, meaning $a_4=a_3=a_2=0$  (see Eq. \eqref{RNradialEOM}-  ). This is the case for:
		\begin{align}
			E^2&=1\,, &   \bar q&= {\rm sign}(E) 1\, , & L^2&= 1-Q^2\, .
		\end{align}
		If we restrict the discussion to positive energies, the ISCO diverges for a charge product $\bar q=1$. This corresponds to the Newtonian case, where the gravitational attraction and electric repulsion of a charged test particle annihilate each other. Values of $\bar q>1$ correspond to a in total repulsive force on a test particle in the Newtonian limit and no bound orbits are possible.
		
		The ISCO further reaches infinity for 
		\begin{align}
			E&\rightarrow 0\,, &   \bar q&= -\infty\, .
		\end{align}
		This can be derived from solving Eqs. \eqref{eq:EISCO-cond} and \eqref{eq:rISCO-cond} for $\bar q$ and $E$ respectively and then calculating the limit $\lim_{r\rightarrow\infty}{(\bar q, E)} $.
		
		In figure \ref{pic:ISCO}(b) Eq. \eqref{eq:rISCO-cond} is plotted implicitly as $r_{\rm ISCO}$ over the charge product $\bar{q}$ for several values of a black hole charge $Q$. 

Here we see that $r_{\rm ISCO}$ grows with increasing $|\bar q|$ in the case of attractive Coulomb interaction $qQ<0$. For repulsive Coulomb interaction, the ISCO radius first decreases to a minimum and then increases again, until it diverges as the charge product approaches one. ISCOs cease to exist for $qQ\geq 1$. In case of an attractive interaction $r_{\rm ISCO}$ grows slower in comparison to the repulsive interaction, but ISCOs exist for all $\bar q<1$. 
		
		For each black hole charge $Q$ one can locate the charge product $\bar q$, for  which the  ISCO reaches its smallest radius. In the case, that the effects of the black hole charge on the spacetime curvature are negligible ($Q\approx 0$), the smallest ISCO is reached in case of an uncharged test particle, and located at $r=6$, according to the discussion above. With increasing $Q$ the radius $r$ of the smallest ISCO decreases. The value of $\bar q$ for which the minimal ISCO is reached is given by 
		\begin{eqnarray}
	 		\label{eQmin}
			\bar q= 2 \sqrt{\frac{-5 Q^2+9\left(1-\sqrt{1-Q^2}\right)}{-9+25 Q^2}} Q\, ,
		\end{eqnarray} 
		 According to Eq. \eqref{eQmin}, the minimal ISCO moves to bigger values of $\bar q>0$ for bigger values of $Q$ and reaches $\bar q=1$  in case of the extreme Reissner-Nordström black hole ($Q=1$). This is at the same time the biggest value of $\bar q$ for which ISCO solutions can be found. The corresponding minimal ISCO lies at 
		\begin{equation}
		r_{\rm ISCO,min,exRN}=3\, . 
		\end{equation}  
		
		\begin{figure}
			\centering
			\includegraphics{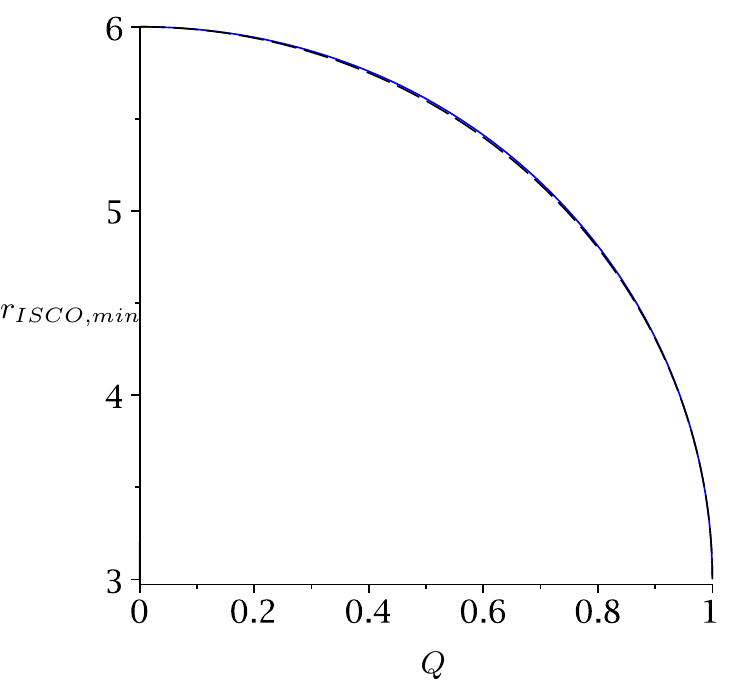}
			\caption{Minimal ISCO radius depending on the black hole charge $Q$ (blue). It shows a nearly circular course, which is fitted by circle $r_{\rm ISCO,min}=3 \left(1+ \sqrt{1-Q^2}\right)$ (black, dashed).  }
			\label{PrISCOminRN}
		\end{figure}
		The course of the minimal ISCO, depending on $Q$ is shown in Fig. \ref{PrISCOminRN}. It is well fitted by a quarter circle given by the Eq.
		\begin{equation}
		r_{\rm ISCO,min}=3 \left(1+ \sqrt{1-Q^2}\right) =3 r_+\, .		
		\end{equation}
	
	\section{ISCO in Kerr-Newman spacetime}
		A discussion of the ISCO behaviour in Kerr-Newman spacetime was already grazed before eg. in \cite{Schroven:2017jsp} for a very small black hole charge. However, we will give a more exhaustive discussion in this chapter.
		 
		The Kerr-Newman metric of a rotating, electrically charged black hole is \cite{Newman:1965my}
		\begin{equation}
		\dd s^2 =  \frac{\rho^2}{\Delta}\dd r^2 + \rho^2\dd\theta^2 + \frac{\sin^2\theta}{\rho^2}\left(\Sigma \dd\phi -a\dd t\right)^2 -\frac{\Delta}{\rho^2}\left(a\sin^2\theta\dd\phi -\dd t\right)^2
		\end{equation}
		with the metric functions
		\begin{align}
		\rho^2&=r^2+a^2\cos^2\theta \nonumber \, ,\\
		\Delta&=r^2-2Mr+a^2+Q^2 \, ,\\
		\Sigma&=r^2+a^2 \nonumber\\
		\end{align}
		and the vector potential $A=\frac{Qr}{\rho^2}(\dd t -a \sin^2\theta\dd\phi)$. The Kerr-Newman black hole has two horizons given by $\Delta=0$
		\begin{equation}
		r_\pm=M\pm\sqrt{M^2-a^2-Q^2} \, .
		\end{equation}
		Therefore horizons exist as long as $M^2>a^2+Q^2$. The ring-like singularity is given by $\rho^2=0$, which is true for $r=0$ and $\theta=\frac{\pi}{2}$.
		
		To derive the equations of motion we use the Hamilton-Jacobi formalism. The Hamilton-Jacobi equation \eqref{eqn:HJD} can be solved with the ansatz
		\begin{equation}
		S= \half\delta\lambda -Et + L\phi + S_r(r) + S_\theta (\theta).
		\end{equation}
		With the help of the Carter constant $K$ the Hamilton-Jacobi equation separates and yields the equations of motion
		\begin{align}
		\label{EOMKNr}
		\left( \frac{\dd r}{\dd\gamma} \right)^2 &= (\bar q^2- \Delta\delta)r^2 -2\bar q r(\Sigma E-La) - K\Delta + (\Sigma E-La)^2 = R(r) \, ,\\
		\left( \frac{\dd \theta}{\dd\gamma} \right)^2 &= K -\delta a^2\cos^2\theta - \frac{(L-aE\sin^2\theta)^2}{\sin^2\theta} \, ,\\
		\left( \frac{\dd \phi}{\dd\gamma} \right) &= \frac{a}{\Delta} (-\bar q r+\Sigma E- La) +\frac{L-aE\sin^2\theta}{\sin^2\theta}  \, ,\\
		\left( \frac{\dd t}{\dd\gamma} \right) &= \frac{\Sigma}{\Delta} (-\bar q r+\Sigma E- La) + a(L-aE\sin^2\theta) \, .
		\label{EOMKNt}
		\end{align}
		We used scaled quantities in the equations of motion
		\begin{equation}
		r\rightarrow \frac{r}{M} \, , \ \lambda\rightarrow \frac{\lambda}{M} \, , \  a\rightarrow \frac{a}{M} \, , \
		Q\rightarrow \frac{Q}{M} \, , \ L\rightarrow \frac{L}{M} \, , \ K\rightarrow \frac{K}{M^2} \, 
		\end{equation}
		and  the Mino time $\gamma$ with $\dd \lambda =\rho^2\dd\gamma$. The equations of motion were solved analytically in \cite{Hackmann:2013pva}.
		
		From the $r$ equation one can define an effective potential by
		\begin{equation}
		R(r)= f(r)(E-V_+)(E-V_-)
		\end{equation}
		so that
		\begin{align}
		V_\pm &= \frac{\bar q r + aL}{\Sigma}\pm\frac{1}{\Sigma}\sqrt{\Delta\left( K +\delta r^2 \right)} \, .
		\end{align}\\
		
		We will analyse the ISCO in the equatorial plane, where the Carter constant is $K=(E-aL)^2=K_{eq}^2$. In the equatorial plane, the coefficients of the polynomial $R=\sum_{i=0}^4 a_i r^i$ are
		\begin{align*}
		a_4 &= E^2-\delta \, , \nonumber\\
		a_3 &= -2E \bar q +2\delta \, , \nonumber\\
		a_2 &= (E^2-\delta) a^2+\bar q^2-\delta Q^2-L^2 \, , \nonumber\\
		a_1 &= 2\bar q a(L-aE)+2(L-aE)^2 \, , \nonumber\\
		a_0 &= -Q^2 (L-aE)^2 \, .
		\label{coeffKN}
		\end{align*}
		As in the previous section, with the help of the rule of Descartes we can deduce conditions for the existence of bound orbits in the and therefore ISCOs
		\begin{align}
		E^2 &< 1 \, ,\\
		\bar q &< 1 \, ,\\
		L^2 &> 1-Q^2 \, , \\
		a &<\sqrt{1-Q^2} \ \text{if} \ L> \sqrt{1-Q^2} \ \text{or} \ a > -\sqrt{1-Q^2} \ \text{if} \ L< -\sqrt{1-Q^2} \, .
		\end{align}
		These conditions apply for bound orbits with $r>r_+$, however, in the Kerr-Newman spacetime bound orbits of charged particles can also exist behind the inner horizon $r<r_-$ or even for negative $r$ \cite{Hackmann:2013pva}.
		
\begin{figure}[h]
			\centering
			\subfigure[]{
				\includegraphics[width=0.45\linewidth]{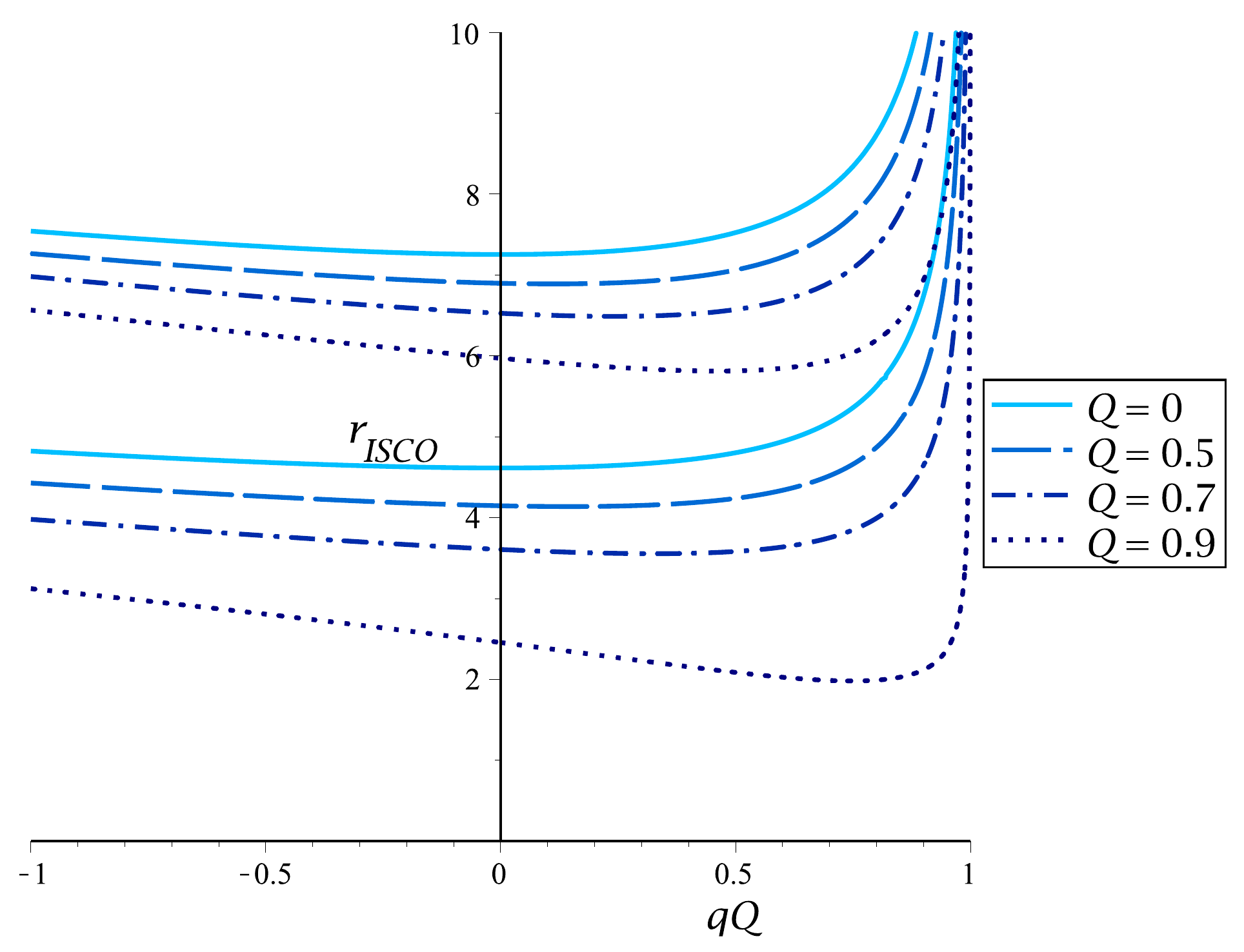}
			}
			\subfigure[]{
				\includegraphics[width=0.45\linewidth]{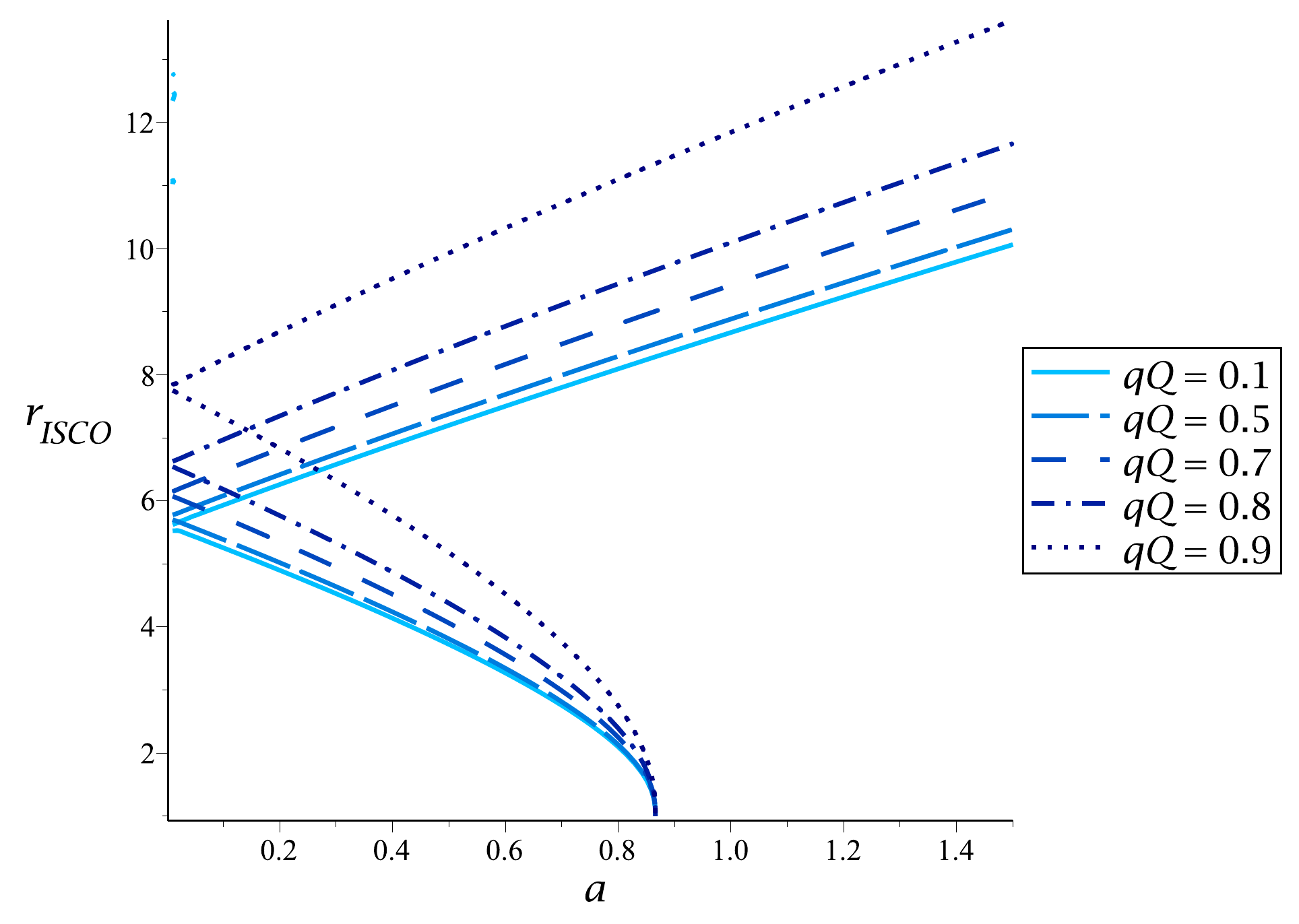}
			}
			\caption{ISCO of electrically charged particles in the Kerr-Newman spacetime. (a) $r_{\rm ISCO}$ is depicted over the charged product $\bar q$ for a black hole spin $a=0.4$ and several values of the black hole charge $Q$. (b)  $r_{\rm ISCO}$ is depicted over black hole spin $a$ for a black hole charge $Q=0.5$ and several values of the charge product $\bar q$. The ISCO diverges when the charge product $\bar q$ reaches one. Corotating ISCOs decrease to smaller radii analogue to Kerr and reach $r=1$ in the extreme case $a=\sqrt{1-Q^2}$. }
			\label{pic:KN-plot}
		\end{figure}

		Using the three conditions for ISCOs \eqref{eqn:cond1}, \eqref{eqn:cond2} and \eqref{eqn:cond3}, we can calculate an equation of the form $f(r, a, Q, q)=0$ which describes the ISCOs. The equation is too long to be displayed here, but we can use it to plot different quantities. Figure \ref{pic:KN-plot} shows the location of the ISCO $r_{\rm ISCO}$ over the charge product $\bar q$ (subfigure (a)) and over the black hole spin $a$ (subfigure (b)). 
		
		The overall course of the ISCO branches is similar to the Reissner-Nordström case. However, outside the horizon, one finds -- analogue to Kerr -- two ISCO solutions. One for the direct, the other one for the retrograde orbit. In case of a not-extreme Kerr-Newman black hole, the ISCO will approach infinity, when the prefactor of the three monomials of $R(r)$ with the three highest degrees in $r$  vanish, meaning $a_4=a_3=a_2=0$  (see Eqs.\eqref{EOMKNr}, \eqref{coeffKN}). This is the case for:
		\begin{align}
			E^2&=1\,, &   \bar q&= {\rm sign}(E) 1\, , & L^2&= 1-Q^2\, .
		\end{align}
		Again, $\bar q=1$ corresponds to the Newtonian case, where the gravitational attraction and electric repulsion of a charged test particle annihilate each other. No bound orbits are possible for $\bar q>1$ and ISCOs cease to exist at this point as well. This can be compared to an in total repulsive force on a test particle in the Newtonian limit.
		
		Analogously to the Reissner-Nordström case the ISCO will grow with increasing $|\bar q|$ for sufficiently big $|\bar q|$. A minimal ISCO therefore exists at some value of $0<\bar q<1$, which moves to to bigger values of $\bar q$ with a growing central charge $Q$. The ISCO minimum occurs for the case of uncharged test particles ($\bar q=0$) if the central charge $Q$ has a negligible effect on the spacetime curvature. An attempt to understand this behaviour was given in section \ref{secI}. This result is independent of the black hole spin $a$, and is derived by verifying that 
		\begin{equation}
			\left. \frac{d}{d\bar q}r_{ISCO}\right|_{\bar q=0}=0\, ,
		\end{equation}
		using Maple.
		
		If the central charge is big enough to significantly effect spacetime, the ISCO of charged particles is smaller than the one in the uncharged case, as long as $|\bar q|$ is sufficiently small. High spins of the black hole do not change this picture, but they strengthen the deviation of the ISCO minimum from $\bar q=0$ for $Q\neq0$.
		
		In Kerr spacetime, the ISCO of the direct orbit approaches $r=1$ for the extreme case ($a=1$). In the case of an extreme Kerr-Newman back hole ($a^2+Q^2=1$), the direct ISCO approaches $r=1$ for
		\begin{align}
		\bar q^*=\frac{1-2a^2}{\sqrt{1-a^2}}<\bar q<1 \, .
			\label{exKNISCIlimit}
		\end{align}
		As mentioned before, no bound orbits are possible for $\bar q>1$. For small $\bar q$, the direct ISCO branch shows the same qualitative shape, as in the non-extreme Kerr-Newman case, but reaches $r=1$ at $\bar q=\frac{1-2a^2}{\sqrt{1-a^2}}$ and stays there for bigger values of $\bar q$. Equation \ref{exKNISCIlimit} is derived by satisfying the ISCO equations and $\frac{\dd^3 }{\dd r^3} R(r)=0\,$ at $r=1$. 
		
		Even though it appears, that the ISCO radius reaches $r=1$ in the extreme case, it will in fact not reach the horizon but actually keeps an infinite distance to it as well as to the photon or marginally bound orbit \cite{Bardeen:1972fi}. The  cause of this deceptive result is the failure of Boyer-Lindquist coordinates to properly resolve the region at the horizon, as the entire section of the spacetime manifold $r<r_{ISCO}$ is projected onto $r=1$.
		
\begin{figure}
			\centering
			\includegraphics{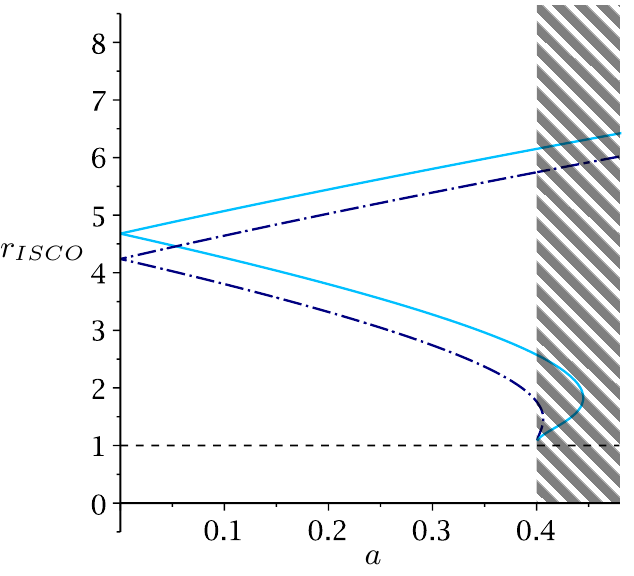}
			\caption{Discussion of the ISCO in Kerr-Newman spacetime for charged particles for black hole charge $Q=\sqrt{1-{0.4}^2}$, and  charge product $\bar q=0.4$, depending on the black hole spin $a$. The black area portrays the region corresponding to overextreme black holes, where naked singularities occur. The extreme black hole case occurs at $a=0.4$. The corotating ISCOs do not reach $r=1$ in this case, since Eq. \eqref{exKNISCIlimit} is not satisfied.}
			\label{ISCOKNfig8}
		\end{figure}

	    The Kerr-Newman ISCO is plotted over the spin $a$ for different values of $Q$ and $\bar q$ in Fig. \ref{pic:KN-plot} (b) and \ref{ISCOKNfig8}.  In Fig. \ref{ISCOKNfig8} $\bar q$ is chosen such, that the condition in Eq. \eqref{exKNISCIlimit} is not satisfied. In contrast to the ISCO branches depicted in Fig. \ref{pic:KN-plot} (b), $r=1$ is not reached for the extremal black hole in this case.
		
		
	\section{Kerr-Sen}	
		The Kerr-Sen spacetime describes a rotating, charged black hole in heterotic string theory. The metric is \cite{Sen:1992ua, Garcia:1995qz}
		\begin{equation}
		\dd s^2 =  \frac{\rho^2}{\Delta}\dd r^2 + \rho^2\dd\theta^2 + \frac{\sin^2\theta}{\rho^2}\left(\Sigma \dd\phi -a\dd t\right)^2 -\frac{\Delta}{\rho^2}\left(a\sin^2\theta\dd\phi -\dd t\right)^2
		\end{equation}
		with the metric functions
		\begin{align}
		\rho^2&=r^2+a^2\cos^2\theta + \frac{Q^2}{M}r \nonumber \, ,\\
		\Delta&=r^2-2Mr+a^2 -\frac{Q^2}{M}r \, ,\\
		\Sigma&=r^2+a^2 + \frac{Q^2}{M}r\nonumber\\
		\end{align}
		and the vector potential $A=\frac{Qr}{\rho^2}(\dd t -a \sin^2\theta\dd\phi)$. The Kerr-Sen black hole has two horizons given by $\Delta=0$
		\begin{equation}
		r_\pm = M+\frac{Q^2}{2M}\pm \sqrt{M^2-a^2+ Q^2+\frac{Q^4}{4M^2}} \, .
		\end{equation}
		The singularity is described by $\rho^2=0$ and depends in contrast to the Kerr-Newman spacetime on the charge. The singularity can have different shapes depending on the charge, see \cite{Flathmann:2015xia}.
		
		In heterotic string theory a dilaton field $\Phi$ is present, which is in this case given by the relation
		\begin{equation}
		\mathrm{e}^{2\Phi} = \frac{r^2+a^2\cos^2\theta}{\rho^2} \, .
		\end{equation}
		The presence of a dilaton field affects the motion of charged particles \cite{Maki:1992up, Pris:1995, Rahaman:2003wv}; then the Hamiltonian is
		\begin{equation}
		\mathcal{H} = \half \mathrm{e}^{-\alpha\Phi}g^{\mu\nu} \left( p_\mu + qA_\mu \right) \left( p_\nu + qA_\nu \right)
		\end{equation}
		where the parameter $\alpha$ is the coupling to the dilaton field. The mass shell condition changes to
		\begin{equation}
		g^{\mu\nu} \left( p_\mu + qA_\mu \right) \left( p_\nu + qA_\nu \right) + \delta\mathrm{e}^{2\alpha\Phi} =0.
		\end{equation}
		$\delta$ describes the mass of the test particle and is $1$ for particles and $0$ for light. To solve the Hamilton-Jacobi equation $\mathcal{H}+\frac{\partial S}{\partial\lambda}=0$ with $p_\mu=\frac{\partial S}{\partial x^\mu}$, we use an ansatz for the action
		\begin{equation}
		S= \half\delta\mathrm{e}^{\alpha\Phi}\lambda -Et + L\phi + S_r(r) + S_\theta (\theta).
		\end{equation}
		Then the Hamilton-Jacobi equation separates in two cases: $\alpha=1$ and $\alpha=0$.
		
		In the case $\alpha=1$ the equations of motion are
		\begin{align}
		\left( \frac{\dd r}{\dd\gamma} \right)^2 &= (\bar q^2- \Delta\delta)r^2 -2\bar q r(\Sigma E-La) - K\Delta + (\Sigma E-La)^2 = R(r) \, ,\\
		\left( \frac{\dd \theta}{\dd\gamma} \right)^2 &= K -\delta a^2\cos^2\theta - \frac{(L-aE\sin^2\theta)^2}{\sin^2\theta} \, ,\\
		\left( \frac{\dd \phi}{\dd\gamma} \right) &= \frac{a}{\Delta} (-\bar q r+\Sigma E- La) +\frac{L-aE\sin^2\theta}{\sin^2\theta}  \, ,\\
		\left( \frac{\dd t}{\dd\gamma} \right) &= \frac{\Sigma}{\Delta} (-\bar q r+\Sigma E- La) + a(L-aE\sin^2\theta) \, .
		\end{align}
		We used scaled quantities as in the Kerr-Newman spacetime and the Mino time $\gamma$ with $\dd \lambda =\mathrm{e}^\Phi\rho^2\dd\gamma$.
		
		In the case $\alpha=0$, i.e. without dilaton coupling, only the $r$-equation is different from the case $\alpha=1$ 
		\begin{align}
		\left( \frac{\dd r}{\dd\gamma} \right)^2 &= R(r) - \delta\Delta Q^2 r = \tilde{R} (r)\, .
		\end{align}
		Again we used scaled quantities and the Mino time as in the Kerr-Newman spacetime. For $\delta=0$ both cases $\alpha=1$ and $\alpha=0$ have the same $r$-equation.
		
		As for Kerr-Newman  one can define an effective potential from the $r$ equation. In the case $\alpha=1$ the effective potential is
		\begin{align}
		V_\pm &= \frac{\bar q r + aL}{\Sigma}\pm\frac{1}{\Sigma}\sqrt{\Delta\left( K +\delta r^2 \right)}
		\end{align}
		and in the case $\alpha=0$
		\begin{align}
		V_\pm &= \frac{\bar q r + aL}{\Sigma}\pm\frac{1}{\Sigma}\sqrt{\Delta\left( K +\delta r^2 + \delta r Q^2\right)} \, .
		\end{align}\\
		
		We will analyse the ISCO in the equatorial plane, where the Carter constant is $K=(E-aL)^2$. In the equatorial plane the coefficients of the polynomial $R=\sum_{i=0}^4 a_i r^i$ are in the case $\alpha =1$
		\begin{align}
		a_4 &= E^2-\delta \, , \nonumber\\
		a_3 &= ( 2E^2+\delta )Q^2 - 2\bar qE + 2\delta \, , \nonumber\\
		a_2 &= (E^2-\delta)a^2 + Q^4 E^2 - 2\bar q Q^2E + \bar q^2 - L^2\, , \nonumber\\
		a_1 &=  2a(\bar q-Q^2E)(L-Ea) + (L-Ea)^2(Q^2+2)\, , \nonumber\\
		a_0 &= 0 \, .
		\end{align}
		In the case $\alpha=0$ the coefficients of the polynomial $\tilde{R} (r)=\sum_{i=0}^4 \tilde{a}_i r^i$ are
		\begin{align}
		\tilde{a}_4 &= E^2-\delta \, , \nonumber\\
		\tilde{a}_3 &= 2E^2 Q^2 - 2E\bar q + 2\delta \, , \nonumber\\
		\tilde{a}_2 &= (E^2-\delta)a^2 + \bar q^2 - 2\bar q Q^2E - L^2 + Q^4E^2 +(Q^2+2)\delta Q^2\, , \nonumber\\
		\tilde{a}_1 &=  2a(\bar q-Q^2 E)(L-Ea) + (L-Ea)^2(Q^2+2) - a^2\delta Q^2\, , \nonumber\\
		\tilde{a}_0 &= 0 \, .
		\end{align}
		
		As in the previous sections, with the help of the rule of Descartes we can deduce conditions for the existence of bound orbits and therefore ISCOs from the $r$-equation. In the case $\alpha=1$ we get the conditions
		\begin{align}
		E^2 &< 1 \, ,\\
		\bar q &< 1 + \frac{3}{2}Q^2 \, ,\\
		L^2 &> \left( 1+\frac{Q^2}{2} \right) ^2 \, ,\\
		a &< 1+\frac{Q^2}{2} \ \text{if} \ L> 1+\frac{Q^2}{2} \ \text{or} \ a > -1-\frac{Q^2}{2} \ \text{if} \ L< -1-\frac{Q^2}{2} \, .
		\end{align}
		and in the case $\alpha=0$ we get
		\begin{align}
		E^2 &< 1 \, ,\\
		\bar q &< 1 + Q^2 \, , \\
		L^2 &> \left( 1 + Q^2 \right)^2 \, ,\\
		a &< 1+\frac{Q^2}{2} \ \text{if} \ L> 1 + Q^2 \ \text{or} \ a > -1-\frac{Q^2}{2} \ \text{if} \ L< -1 - Q^2 \, .
		\end{align}
		These conditions apply for bound orbits with $r>r_+$, however, in the Kerr-Sen spacetime bound orbits of charged particles can also exist behind the inner horizon $r<r_-$ or even for negative $r$.\\
		
\begin{figure}[h]
			\centering
			\subfigure[]{
				\includegraphics[width=0.45\linewidth]{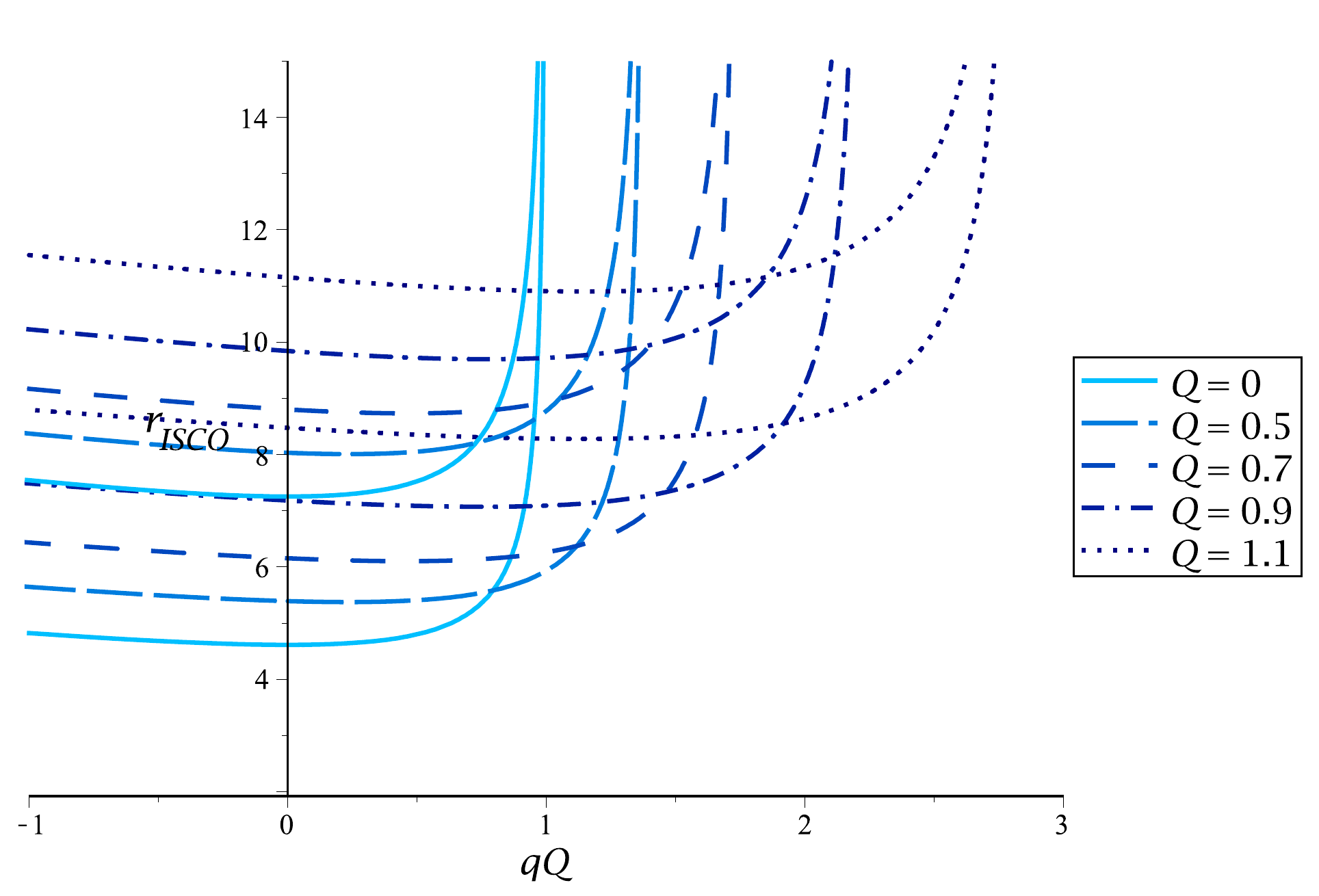}
			}
			\subfigure[]{
				\includegraphics[width=0.45\linewidth]{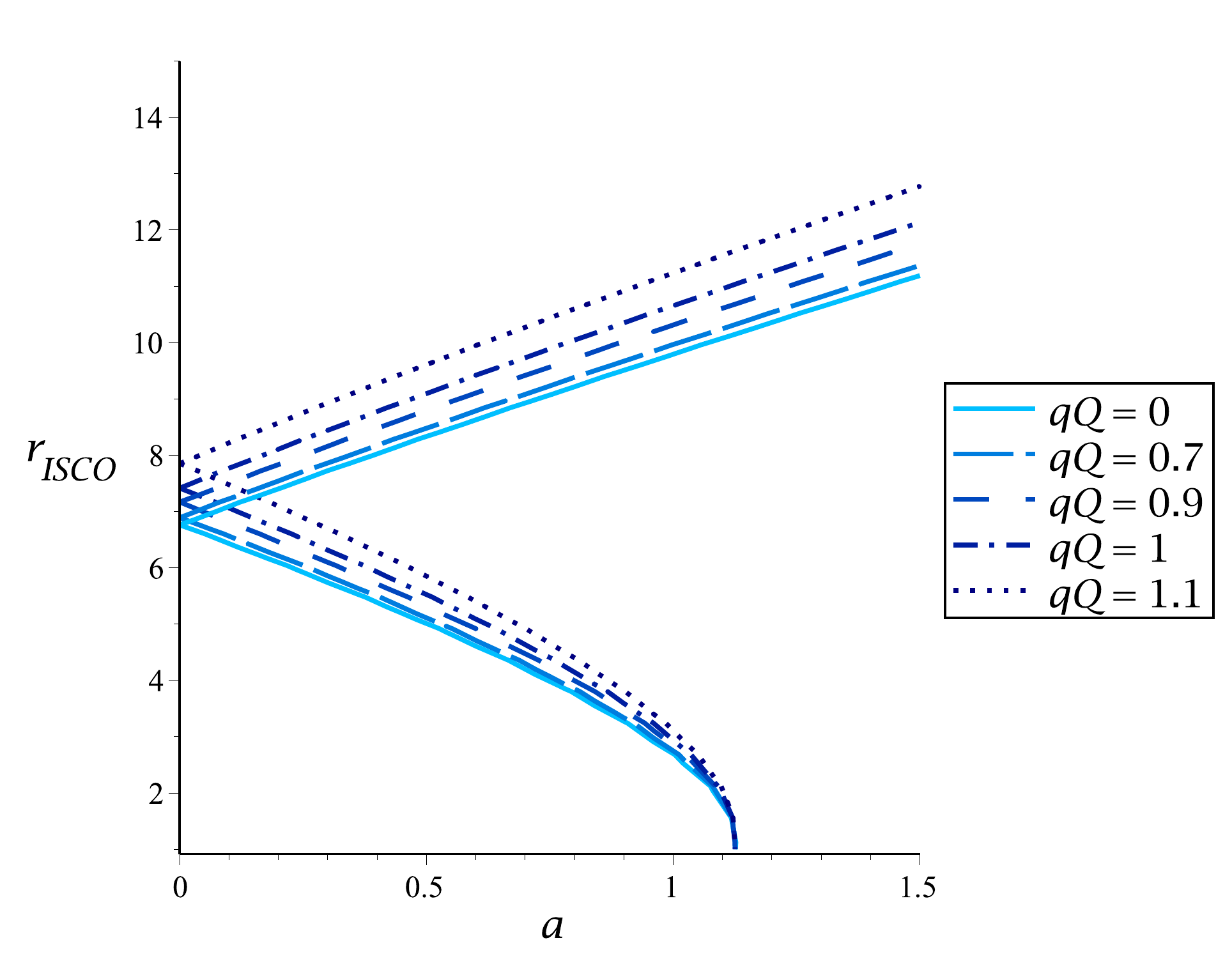}
			}
			\subfigure[]{
				\includegraphics[width=0.45\linewidth]{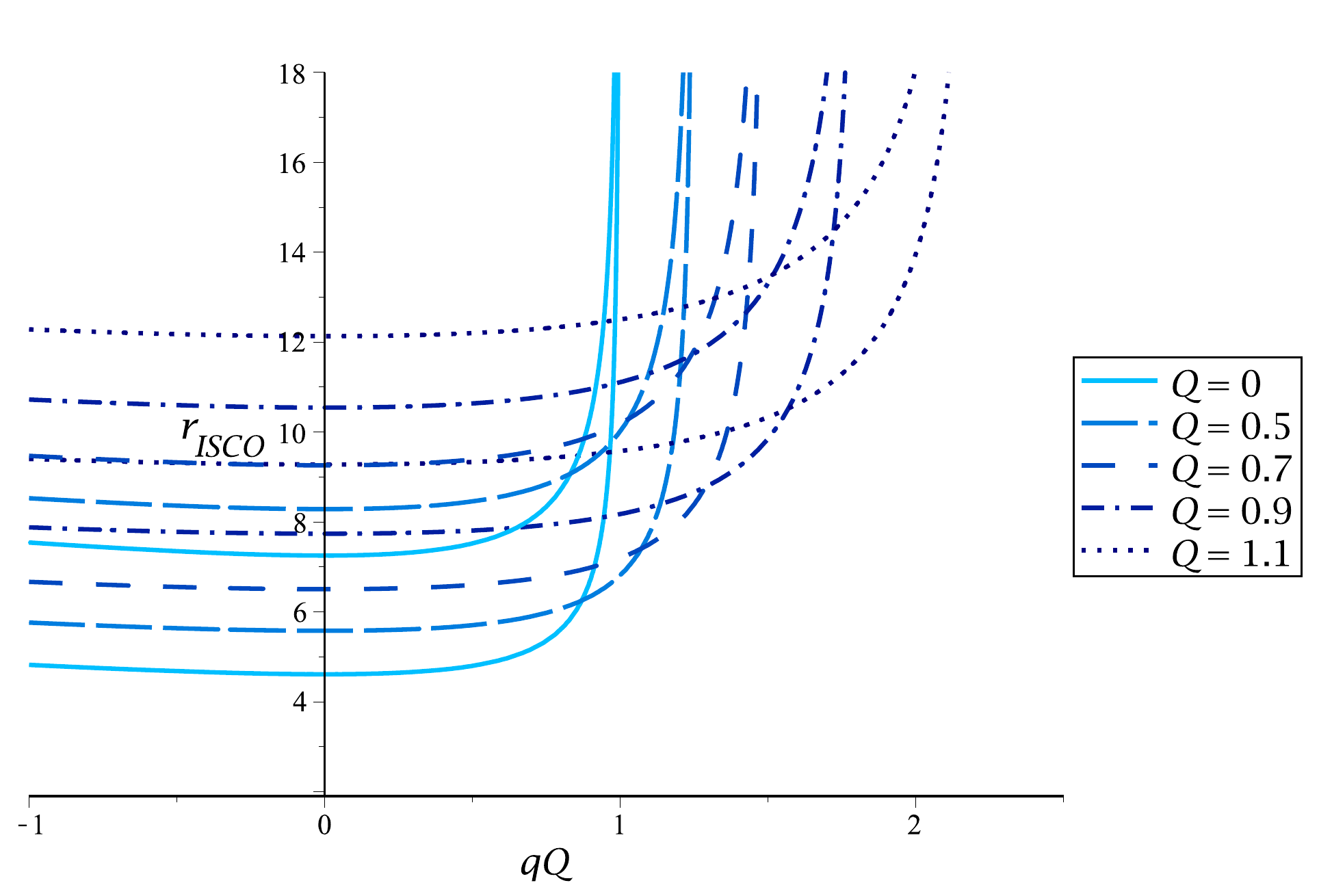}
			}
			\subfigure[]{
				\includegraphics[width=0.45\linewidth]{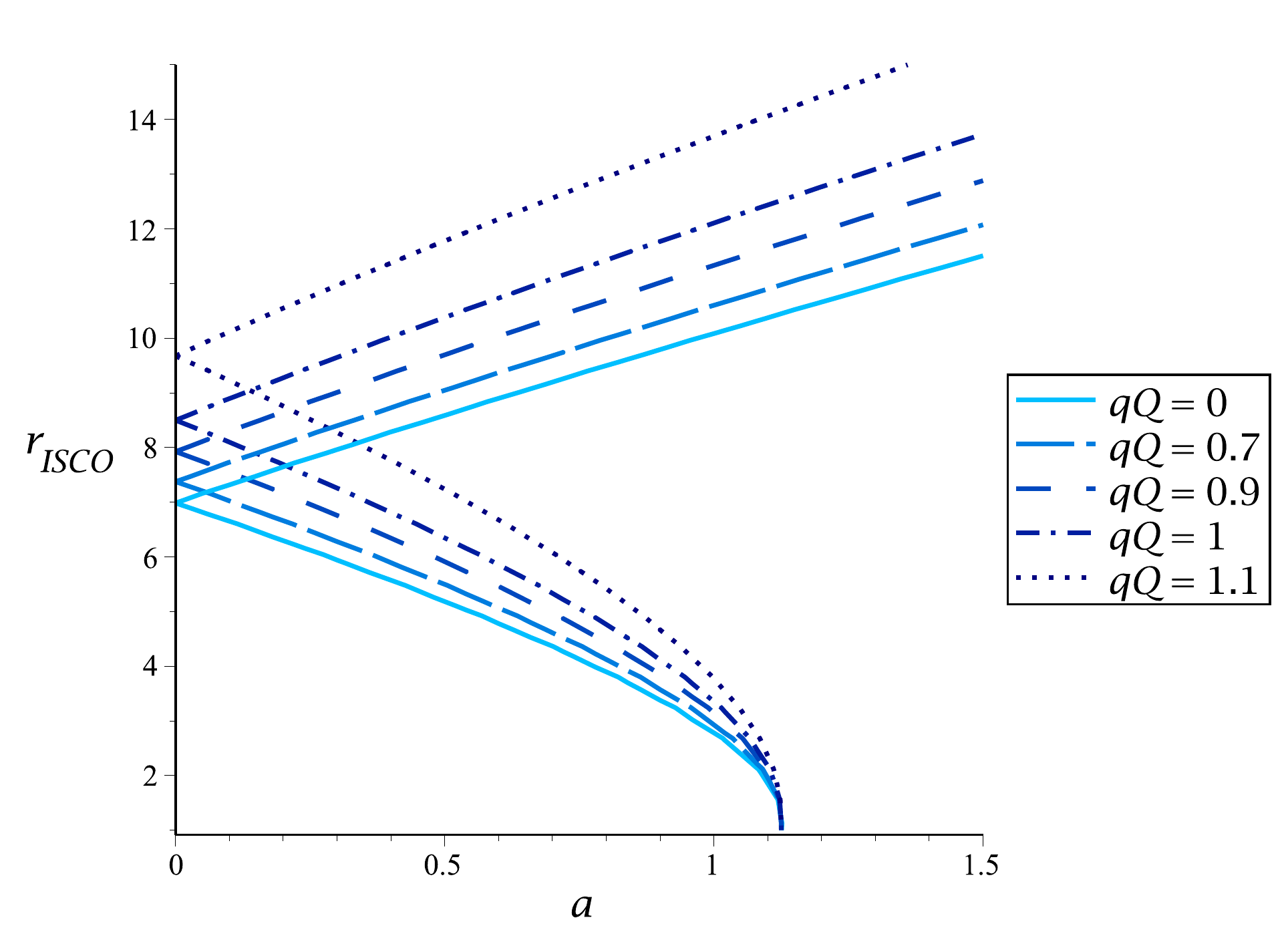}
			}
			\caption{ISCO of electrically charged particles in the Kerr-Sen spacetime. (a) $r_{\rm ISCO}$ over $qQ$ for $\alpha=1$, $a=0.4$ and several values of $Q$. (b)  $r_{\rm ISCO}$ over $a$ for $\alpha=1$, $Q=0.5$ and several values of $qQ$. (c) $r_{\rm ISCO}$ over $qQ$ for $\alpha=0$, $a=0.4$ and several values of $Q$. (d)  $r_{\rm ISCO}$ over $a$ for $\alpha=0$, $Q=0.5$ and several values of $qQ$.}
			\label{pic:KS-plot}
		\end{figure}

		Using the three conditions for ISCOs \eqref{eqn:cond1}, \eqref{eqn:cond2} and \eqref{eqn:cond3}, we can calculate an equation of the form $f(r, a, Q, q)=0$ which describes the ISCOs. The equation is too long to be displayed here, but we can use it to plot different quantities. Figure \ref{pic:KS-plot} shows the ISCO in the Kerr-Sen spacetime. The case $\alpha=1$ is depicted in figure \ref{pic:KS-plot}(a) and (b), (a) shows the location of the ISCO $r_{\rm ISCO}$ over $\bar q$ and (b) shows $r_{\rm ISCO}$ over $a$. The case $\alpha=0$ is depicted in figure \ref{pic:KS-plot}(c) and (d), (c) shows the location of the ISCO $r_{\rm ISCO}$ over $\bar q$ and (d) shows $r_{\rm ISCO}$ over $a$.
		Due to the rotation of the Kerr-Sen black hole we get two ISCO solutions, the upper branch describes the counter-rotating ISCO, which is further away from the black hole than the co-rotating ISCO at the lower branch. Overall the behaviour at both branches is similar to the Reissner-Nordström black hole. $r_{\rm ISCO}$ grows with increasing $|\bar q|$ in the case of attractive Coulomb interaction $qQ<0$. For repulsive Coulomb interaction, the ISCO radius first decreases to a minimum and then increases again, until it diverges. ISCOs cease to exist for $\bar q\geq 1 + \frac{3}{2}Q^2$ in the case $\alpha =1$ and for $\bar q\geq 1+ Q^2$ in the case $\alpha =0$. In case of an attractive interaction $r_{\rm ISCO}$ grows slower in comparison to the repulsive interaction, but ISCOs exist for all $\bar q<1$.
		
		Interestingly, for $\alpha=0$ the minimal ISCO is found in the range $qQ \in \left[0,1+Q^2\right[$, however, for $\alpha=1$ the minimal ISCO is at $qQ=0$ for all $Q$. 
		
		Note that in the Reissner-Nordström and in the Kerr-Newman spacetime ISCOs of charged particles cease to exist for  $\bar q\geq 1$. In the Kerr-Sen spacetime this upper limit for ISCOs depends on the charge of the black hole and is shifted to larger $\bar q$ for growing $Q$.
		
		Furthermore, if the charge $Q$ of the black hole increases the behaviour of the ISCO in the Kerr-Sen is different from the Reissner-Nordström black hole and the Kerr-Newman black hole. In the Reissner-Nordström or Kerr-Newman spacetime, the position $r_{\rm ISCO}$ of the ISCO is closer to the black hole for increasing charge $Q$. In the Kerr-Sen spacetime however, the position  $r_{\rm ISCO}$ of the ISCO is further away from the black hole if the charge $Q$ of the black hole increases.

	\section{Outermost stable circular orbits found beyond the horizon}
		A complete picture of the ISCO discussion requires to cast a glance on the orbits beyond the inner horizon. Negative radii cannot be reached in Reissner-Nordström spacetime. The non-rotating, charged black hole possesses a spacetime singularity at $r=0$. All particle trajectories have to terminate there, and are not allowed to reach negative radii. It is still worth taking a look, especially with regards to a comparison with Kerr-Newman spacetime. In the rotating counterpart of the Reissner-Nordström black hole, a ring singularity allows a transition from positive to negative radii. 
		
		The radial equation of motion in Kerr-Newman and Reissner-Nordström spacetime is a polynomial of order 4. According to Eq. \eqref{EOMKNr} $R(r)$ has to be positive between the horizons.  On the other hand it has to be negative at $r=0$. 
		
\begin{figure}
			\centering
			\includegraphics[scale=0.7]{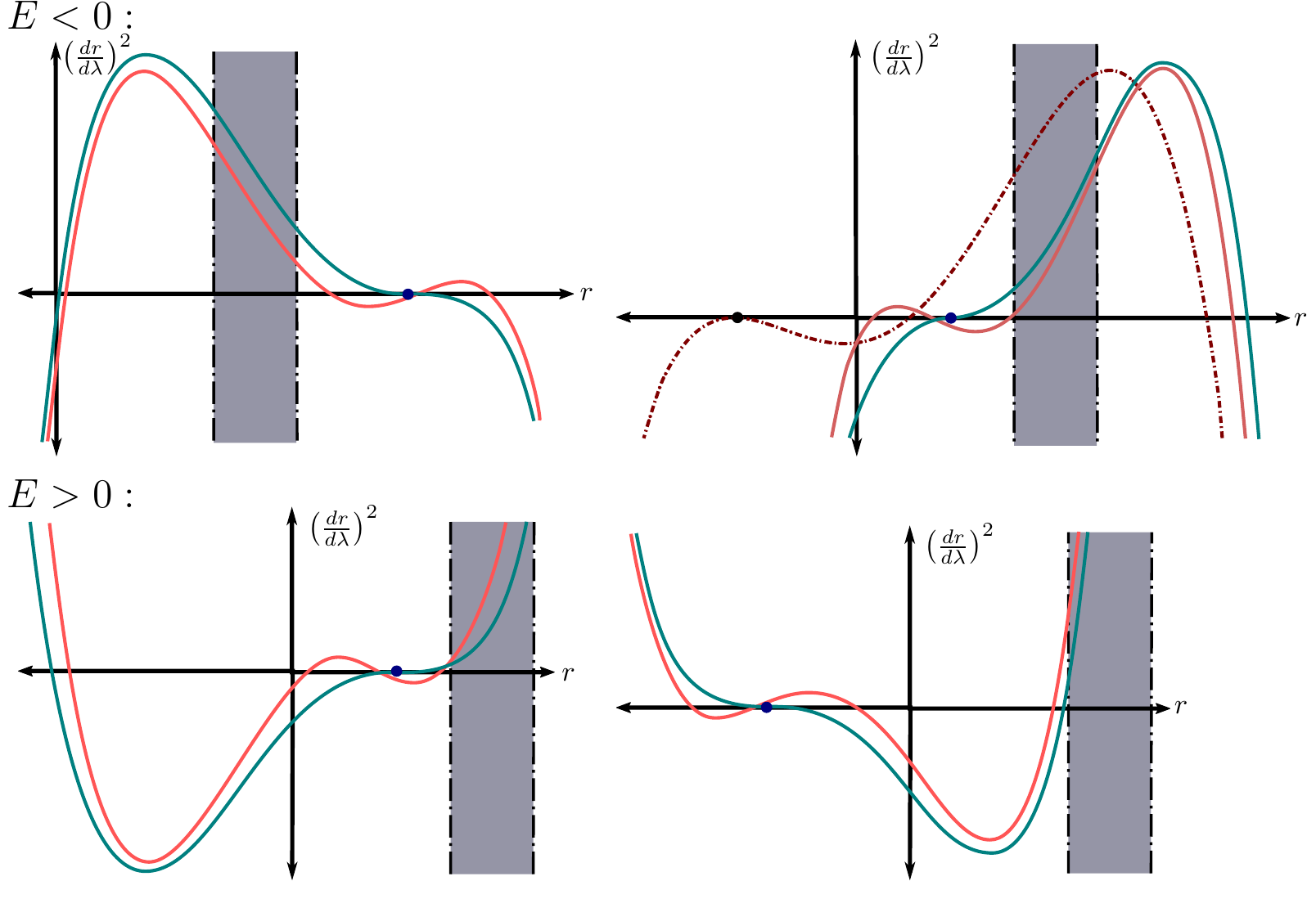}
			\caption{Four qualitatively different possible potentials for the radial motion.}
			\label{potfig1}
		\end{figure}

		Keeping these properties in mind, four qualitatively different configurations can be found for $R(r)$ (see Fig. \ref{potfig1}). Depending on the sign of $(E^2-1)$, circular stable (local maxima) and unstable orbits (local minima) are possible for positive radii bigger than the outer horizon, smaller than the inner horizon, or negative radii. It becomes clear from the plots in Fig. \ref{potfig1}), that a radius $r_{\mathrm{III}}$ satisfying Eqs. \eqref{eqn:cond1}-\eqref{eqn:cond3} represents the ISCO for $r>r_+$, but a -- so to say -- outermost stable circular orbit (OSCO) for radii $0<r<r_-$. In the case of negative radii, $r_{\mathrm{III}}$ represents -- again -- an OSCO, in that sense, that all $\left|r\right|<\left|r_{\mathrm{III}}\right|$ are stable, but all $\left|r\right|>\left|r_{\mathrm{III}}\right|$ are unstable. It seems, that behind the horizon one might find a region of stable circular orbits around $r=0$ for certain particle and black hole charges $q$ and $Q$. 
	
		\begin{figure}
			\centering
			\includegraphics[scale=0.75]{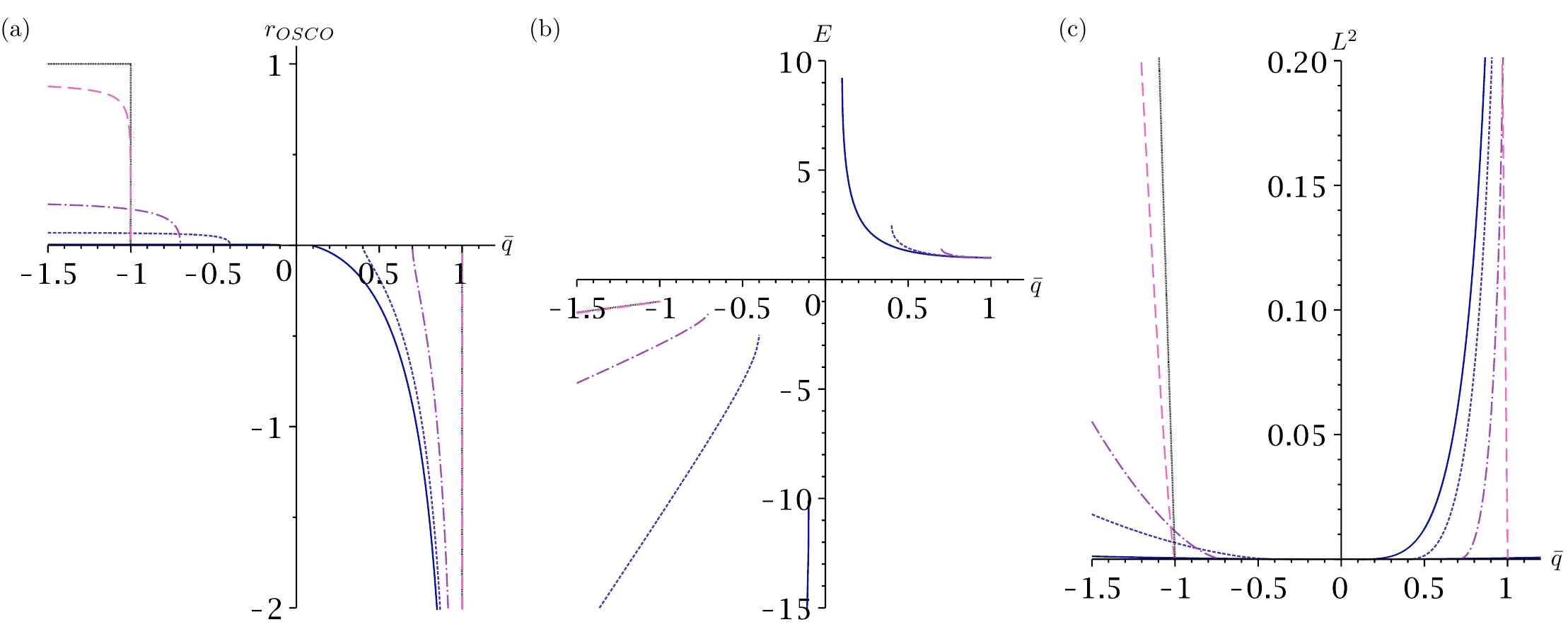}
			\caption{Solutions for an outermost stable circular orbit  (OSCO) in Reissner-Nordström spacetime behind the inner horizon $r<r_-$ for different values of $Q$: $Q=0.1$ (blue, solid),$Q=0.4$ (dark violet, short-dashed), $Q=0.7$ (violet, dash-dotted line) $Q=0.999$ (bright violet, long-dashed), and $Q=1$ (black, thin-dashed). (a) shows the OSCO radius, (b) shows the corresponding energy and (c) the corresponding squared angular momentum. Negative energies occur for positive OSCOs $<r_-$.}
			\label{ISCObhRN}
		\end{figure}

		In the Reissner-Nordström limit bound orbits are found behind the horizon \cite{Grunau:2010gd}. In agreement with this, two branches of OSCOs are found for radii smaller than the inner horizon.  Due to the symmetries in the equations of motion \eqref{EOMKNr}-\eqref{EOMKNt} (set $a=0$ for the Reissner-Nordström limit), we reduce the discussion to solutions with a positive time evolution ($dt/d\gamma>0$). OSCOs with a negative time evolution are found at the same radii, but ($\bar q\rightarrow -\bar q, E\rightarrow -E$).
		
		The OSCO branches and their energies and angular momenta are plotted in Fig. \ref{ISCObhRN} for different values of $Q$. One branch occurs for negative radii and a positive charge product $\bar q$, and one for positive radii smaller than the inner horizon and a negative charge product $\bar q$. The two branches for $r<r_-$ "merge" at $r=0$ for $\left|\bar q\right|=\bar q_{\rm ST}$. This point can be understood as a germ for the two OSCO solution branches. It satisfies not only to the conditions for an ISCO or OSCO, but
		\begin{equation}
		\label{STcond}
		\frac{\dd^k }{\dd r^k} R(r)=0\, , \text{ for } k=0..3\, ,
		\end{equation}
		and is located at
		\begin{align}
		E&=\frac{1}{\bar q}\, , & \bar q^2&=a^2+Q^2\eqqcolon \bar q_{\rm ST}^2\, , & K_{eq}&=0\, , & r&=0\, ,
		\label{swtail}
		\end{align}
		in the general Kerr-Newman case. The conditions in \eqref{STcond} determine a germ of a Swallowtail catastrophe for $\mathbf{R}(r)=\int R(r)\,dr$.
		 
		The swallowtail point $\bar q_{\rm ST}$ marks the absolute value $|\bar q|$, above which stable orbits exist for both positive and negative radii $r<r_-$. The OSCO solution for negative radii diverges at $\bar q=1$, meaning that stable circular orbits can be found for all $r<0$, if $\bar q>1$. On the other hand no stable circular orbits are found for  $r<0$ and $|\bar q|< |\bar q|_{\rm ST}$. According to Eqs. \eqref{swtail} $\bar q_{\rm ST}$ moves to $|\bar q|=1$ for bigger values of $Q$ and reaches $|\bar q|=1$ for the extremal case $Q=1$. The OSCO solution for positive radii behind the horizon approaches $r_{\rm OSCO}=1$. This leads to a step function
		\begin{equation}
			r_{\rm OSCO}=\left\{
								 \begin{array}{c}
									1 \text{ for } eQ<-1 \\
									0 \text{ for } eQ =-1 \\
							     \end{array}		
						 \right.
		 \label{OSCOsf}
		\end{equation} 
		for the OSCO solutions in the extreme case. The more the values of $Q$ approaches the extreme case $Q=1$, its two OSCO solution branches approach the step function for positive radii and the course $\bar q=1$ for negative radii (Fig. \ref{ISCObhRN}).
		
		\begin{figure}
			\centering
			\includegraphics[scale=0.95]{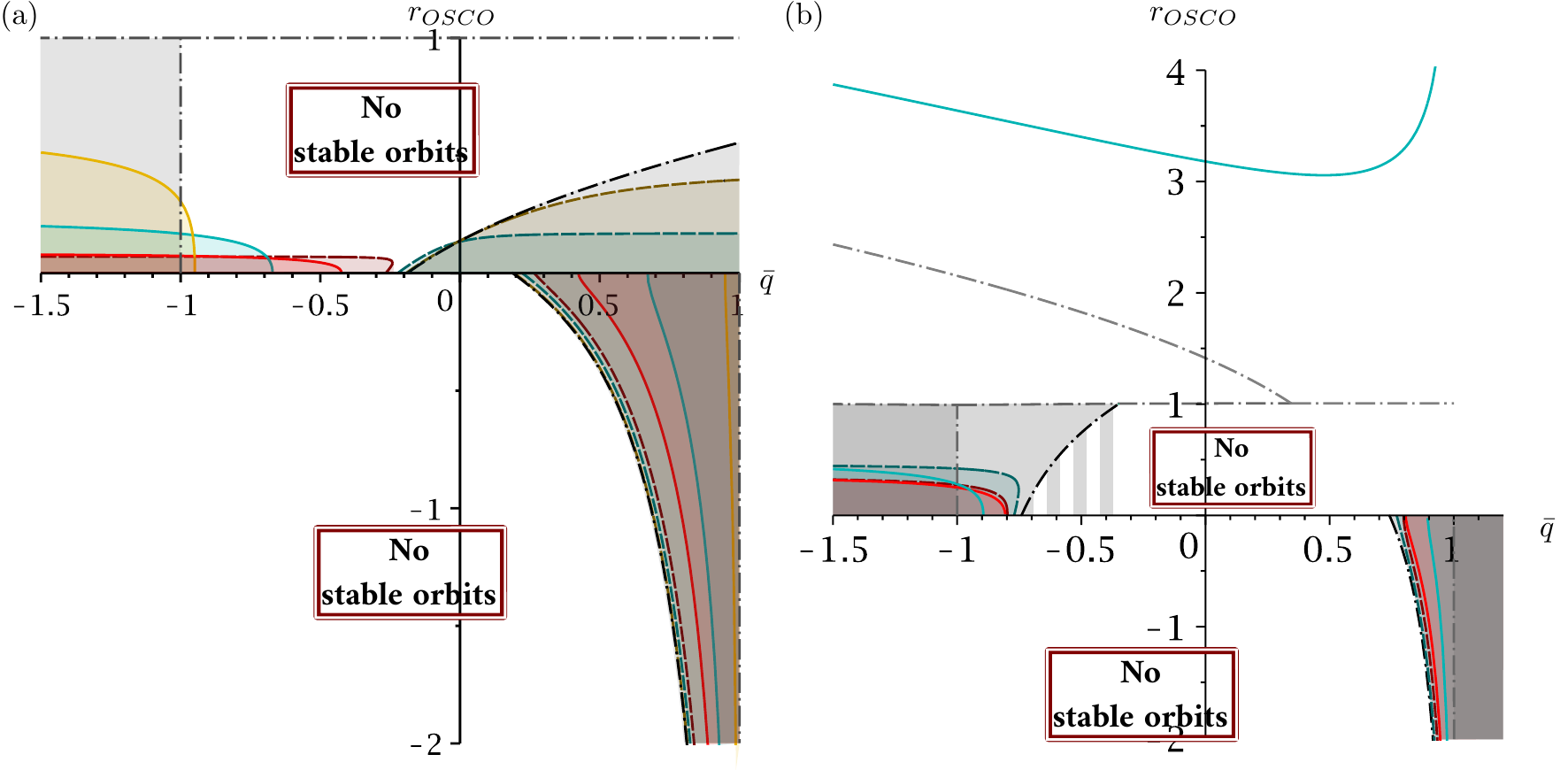}
			\caption{OSCO solutions in Kerr-Newman spacetime for two different values of $Q$ and different spins. (a) OSCO solutions for $Q=0.3$ and four different values of the spin: $a=0.3$ (red),$a=0.6$ (blue), and $a=0.9$ (yellow) and the extreme black hole case $a=\sqrt{1-Q^2}$ (black).  (b) OSCO solutions for $Q=0.8$ and three different values of the spin: $a=0.1$ (red),$a=0.4$ (blue), and the extreme black hole case $a=0.6$ (black).
			"Counterrotating" OSCOs are plotted as dashed lines and in a darker tone. For negative radii, the OSCO diverges to $-\infty$ at $\bar q=1$. The extreme black hole case is plotted as a dash-dotted line. Like in the Reissner-Nordström case, the corotating OSCO branch approaches the course of the step function as given Eq. \eqref{OSCOsf} for $r>0$ and the course $\bar q=1$ for negative radius. The regions in $[r,\bar q]$ for which stable orbits can be found is coloured in the respective color of the corresponding $a$.} 
			\label{OSCOKN}
		\end{figure}
		In the Kerr-Newman case bound solutions were found behind the horizon by \cite{Hackmann:2013pva}. Due to the symmetries in the equations of motion \eqref{EOMKNr}-\eqref{EOMKNt}, we will restrict our discussion to branches of solutions, that show a positive time evolution for radii not too close to the singularity $r=0$. A second set of solutions is found with the same radii, but ($\bar q\rightarrow -\bar q, E\rightarrow -E, L=-L$). In comparison to the Reissner-Nordström limit a second branch of solution occurs for both areas: $\left[0<r<r_-\right]$ and $\left[r<0\right]$ (see Fig. \ref{OSCOKN}). The set of branches that shows the same qualitative behaviour as in the Reissner-Nordström case corresponds to corotating orbits. Like in the Reissner-Nordström case, the two branches "merge" at the location of the Swallowtail germ given in  Eq. \eqref{swtail}. Stable corotating orbits can be found for $\bar q>|\bar q|_{\rm ST}$ for negative radii and for  $\bar q<-|\bar q|_{\rm ST}$ in case of positive radii. 
		
		In the extreme black hole case ($a^2+Q^2=1$), the corotating OSCO branch for positive radii approaches the step function \eqref{OSCOsf} like in the nonrotating limit. The corotating OSCO branch for negative radii approaches the course $\bar q=1$. 
		
		The second set of OSCO branches shows the property $K_{eq}=L-aE<0$. So one might associate these OSCO branches with "counter-rotating" orbits. However, since negative energies are possible behind the horizon, $K_{eq}<0$ might not necessarily correspond to an actual negative angular momentum of the particle for the whole branch of solutions.
		
		The "$K_{eq}<0$"-OSCO branches show the same qualitative behaviour as the corotating OSCOs for negative radii. They start at smaller charge products $\bar q>0$, compared to their corotating counterparts, but also diverge for $\bar q=1$. One branch of "$K_{eq}<0$"-OSCOs lies at $r=1$ for an extreme black hole and runs from $-\infty$ to a maximal $\bar q=-\bar q^*$ (given in Eq. \eqref{exKNISCIlimit}, where the corotating ISCO branch reaches $r=1$), if $\bar q^*>0$ (see Fig. \ref{OSCOKN} (a) ). Otherwise --if $\bar q^*<0$-- it starts at $-\bar q^*$, and decreases for smaller $\bar q$, until the OSCO reaches $r=0$ (see Fig. \ref{OSCOKN} (b) ). The "$K_{eq}<0$"-OSCO branches of non-extreme black holes will in general follow the same course.

\section{Conclusion}
	The existence of an innermost stable circular orbit for test particle motion around a compact object is a purely relativistic phenomenon. It is therefore likely, that any  intuitive expectations one might have on its behaviour turn out to be wrong. 
	In this paper we discussed the -- at times counter intuitive -- behaviour of ISCOs for charged particles in Reissner-Nordström, Kerr-Newman, and Kerr-Sen spacetime.
	
	A minimal ISCO occurs in the Reissner-Nordström spacetime and in the Kerr-Newman spacetime for co- and counterrotating orbits at a particle-black hole charge product $qQ$ in the range $qQ \in \left[0,1\right[$. In other words, the ISCO location is pushed further outwards for both -- attractive and repulsive -- electromagnetic interactions between the black hole and the particle, above a certain value of $|qQ|$. All ISCO solutions in Reissner-Nordström and Kerr-Newman spacetime  diverge to infinity at $qQ=1$ and cease to exist anywhere above $qQ>1$. A too strong repulsive electromagnetic interaction prohibits any stable circular orbits in this case. 
	
In the Kerr-Sen spacetime one needs to take into account the dilaton coupling of the test particles. In this article we concentrated on the cases $\alpha=0$ and $\alpha=1$, where the Hamilton-Jacobi equation separates and yields equations of motion. For $\alpha=0$ the minimal ISCO is found in the range $qQ \in \left[0,1+Q^2\right[$. For $\alpha=1$ however, the minimal ISCO stays at $qQ=0$ for all $Q$. As in Reissner-Nordström and Kerr-Newman spacetime, the ISCO radius increases both for attractive and repulsive electromagnetic interaction above a certain value of $|qQ|$. The ISCO radius diverges for $qQ=1+Q^2$ if $\alpha=0$ and for $qQ=1+\frac{3}{2}Q^2$ if $\alpha=1$. After these values ISCOs cease to exist.
	
	The effect of even quite strong electromagnetic fields on the spacetime curvature is in general very small and can be neglected in most scenarios. In this case, non-charged particles have the smallest possible ISCO and the ISCO will increase both for growing attractive and repulsive electric forces on the test particle.
	
For a rising effect of the black hole charge on the spacetime curvature, the minimal ISCO moves from $qQ=0$ to increasing values of the charge product $qQ$ (except for particles with $\alpha=1$ in the Kerr-Sen spacetime, where the minimal ISCO stays at $qQ=0$). In Reissner-Nordström spacetime the minimal ISCO moves up to $qQ\rightarrow 1$ for an extremal black hole ($Q=1$). Also the minimal ISCO radius decreases for increasing charge $Q$ of an Reissner-Nordström or Kerr-Newman black hole. The closest distance to the black hole is $r_{\rm ISCO}=3$ for an extremal Reissner-Nordström black hole. Bearing that in mind one can conclude, that, if the total mass and spin of a black hole are known, an ISCO smaller than the one expected for Kerr or Schwarzschild indicates a charge strong enough to significantly affect spacetime curvature.

The ISCO in the Kerr-Sen spacetime behaves differently here. For increasing charge $Q$ the ISCO radius will be further away from the black hole.

	In the uncharged case of an extremal spinning Kerr black hole ($a=1$), the  corotating ISCO approaches the horizon at $r=1$. The same behaviour is found for an extremal Kerr-Newman black hole ($a^2+Q^2=1$) -- but only for a certain range of particle-black hole charge products. If the effect of the black hole charge on the spacetime curvature is negligible ($Q\approx 0$, $a\approx 1$), the corotating ISCO approaches the horizon for all $qQ<1$. The range of $qQ$, for which this behaviour can be found, shrinks with rising $Q$, and finally vanishes for an extremal charged, nonrotating black hole ($Q\rightarrow 1, a\rightarrow 0$).  
	
	We also studied the region behind the horizon in Reissner-Nordström and Kerr-Newman spacetime. Here outermost stable circular orbits (OSCOs) are found instead of ISCOs. They exist for positive and negative radii and border an area around the curvature singularity, in which stable circular orbits are possible. The existence and size of this region depends on how much the black hole charge $Q$ affects spacetime curvature, and on the charge product $qQ$. However, the discussion of OSCOs behind the horizon gets complicated quickly in case of the charged, spinning black hole, which limits the number of final conclusions that can be drawn about this region.
\\
	
	For future research it might be interesting to study the influence of magnetic charge on the radius of the ISCO and consider magnetically as well as electrically charge particles in black hole spacetimes with magnetic and electric charges.

\section{Acknowledgement}
	We would like to thank Jutta Kunz and Vladimír Karas for fruitful
	discussions. SG gratefully acknowledges support by the DFG (Deutsche
	Forschungsgemeinschaft/German Research Foundation) within the Research
	Training Group 1620 ``Models of Gravity''. 
	KS gratefully acknowledges support by the Czech Science Foundation collaboration project (GAČR 19-01137J).

		\bibliographystyle{unsrt}
		\bibliography{references}

\end{document}